\documentclass[a4paper,12pt]{article}

\usepackage{amsmath,amssymb,amsfonts}
\usepackage[dvips]{graphicx}
\usepackage{epsfig}

\usepackage{cite}
\usepackage{amsmath,amssymb,amsfonts,graphicx}
\usepackage{epsfig}
\usepackage[left=3cm,top=3.5cm,right=3cm,bottom=3.5cm,bindingoffset=0cm]{geometry}

\newcommand {\beq} {\begin{equation}}
\newcommand {\eeq} {\end{equation}}
\newcommand {\beqn}{\begin{eqnarray}}
\newcommand {\eeqn} {\end{eqnarray}}
\newcommand{\bea}{\begin{eqnarray}}
\newcommand{\eea}{\end{eqnarray}}

\def\tg{\widetilde{g}}
\def\p{\varphi}
\def\tp{\widetilde{\varphi}}

\def\1{\mathbbm{1}}

\def\MP{M_{P}}

\def\p{\varphi}
\def\GN{G_{N}}
\def\tMP{\widetilde{M}_{ P}}
\def\tGN{\widetilde{G}_{ N}}

\def\tnabla{\widetilde{\nabla}}
\def\tf{\widetilde{f}}
\def\S{\mathbb{S}}
\def\tc{\widetilde{c}}
\def\tS{\widetilde{S}}

\def\tm{\widetilde{m}}
\def\ta{\widetilde{a}}
\def\th{\widetilde{h}}
\def\parlarge{\phantom{\frac{1}{2}}}
\def\tA{\widetilde{A}}
\def\tF{\widetilde{F}}
\def\ts{\widetilde{\sigma}}

\numberwithin{equation}{section}

\begin{document}

\title{
\begin{flushright}\ \vskip -1.5cm {\small {IFUP-TH-2015}}\end{flushright}
\vskip 40pt
\bf{ \Large Born Reciprocity and Cosmic Accelerations }
\vskip 20pt}
\author{S. Bolognesi\\[20pt]
{\em \normalsize
Department of Physics â E. Fermiâ University of Pisa and INFN, Sezione di Pisa}\\[0pt]
{\em \normalsize
Largo Pontecorvo, 3, Ed. C, 56127 Pisa, Italy}\\[3pt]
{ \normalsize Email: stefanobolo@gmail.com}}
\vskip 10pt
\date{August 2015}
\maketitle
\vskip 0pt

\begin{abstract}

The trans-Planckian theory  \cite{Bolognesi:2012fq} is a model that realizes concretely the Born reciprocity idea, which is the postulate of absolute equivalence between coordinates $x$ and momenta $p$. 
This model is intrinsically global, and thus it is naturally implemented in a cosmological setting.  
Cosmology and Born reciprocity are made for each other.
Inflation provides  the essential mechanism to suppress the terms coming from the dual part of the action.
The trans-Planckian theory provides an explanation for  the  present acceleration of the universe scale factor. 
This is possible just considering a simple model that contains gravity, one gauge field  plus one matter field (to be identified with dark matter), together with the  reciprocity principle.

\end{abstract}
\newpage

\section{Introduction}

In the uncertainty principle,   coordinates   $x_{\mu}$ and momenta $p_{\mu}$ enter in a  symmetrical way. 
Once dynamics is taken into account, this duality is ruined. 
Born \cite{born} was the first to envision the possibility that in the fundamental theory this duality should instead be manifest.
The existence of this duality requires a new fundamental energy scale that we call $M$. 
At energy scales much lower than $M$, we should recover the ordinary theories with an asymmetry between coordinates  and momenta. 
We now apply this idea to  gravitational problems and thus $M$ is taken to be of order of the Planck scale \cite{Bolognesi:2012fq}; for this reason, we use the denominations ``Born reciprocity'' and ``trans-Planckian duality'' indistinguishably. For other recent approaches to Born reciprocity in relation to gravity see  \cite{grav}.

The simplest theory that exhibits this duality is the so-called relativistic harmonic oscillator \cite{rho}:
\beq
\label{rhoaction}
{\mathbb S} = \int d^4 x \left( \partial_{\mu} \p^* \partial^{\mu} \p +  M^4 x_{\mu} x^{\mu} \p^* \p \right) \ .
\eeq
 If all the particles were subject to the potential term of the previous action, we would have a tiny universe, bounded at the length scale $M^{-1}$.
This is the first obstacle to overcome in realizing Born reciprocity at the Planck scale, is that the the free kinetic term  would dominate the action for energies much higher than $M$ and not, as we would like instead, much lower than $M$.
The other, and more technical, problem is how to generalize the action of the relativistic harmonic oscillator to the cases in which we have gravity,  gauge interactions, and a generic set of matter fields and interactions. 
For this, we introduce a generalized version of the Fourier transform.

We implement the theory in a cosmological setting for a Friedman-Robertson-Walker (FRW) metric. 
The generalized Fourier transform requires certain boundary conditions to be satisfied. 
We will show that an  early stage inflationary expansion period \cite{inflation} can  make the dual sector of the theory unobservable at low energies. 
We explain how the periods of accelerated expansion of the universe, both inflation and the current period dominated by dark energy, can be explained within this formalism. The main novelties of this paper with respect to the previous ones \cite{Bolognesi:2012fq} are the discussion of how  to explain the inflationary period using this duality and a simpler approach to the boundary conditions problem.

There are many strategies to address the dark energy problem \cite{darkenergy}. We can divide them into two broad categories: the ones in which  dark energy is due to a fundamental cosmological constant $\Lambda_{ \rm fund}$, and the others  in which $\Lambda_{ \rm fund}$ is zero and dark energy has, instead, another dynamical origin.    Our approach belongs to the second category.  Dark matter has a modified equation of state at late-time, when the effect of the dual terms of the action becomes relevant, and this effect induces a distortion of its equation of motion and, thus, an effective  cosmological constant contribution to the energy-momentum tensor. This effect is compatible with the observed value of the present value of the universe acceleration if the inflationary stage lasts rougly  the  amount of time required to solve the horizon problem.


The paper is organized as follows. In Section \ref{theory} we introduce the trans-Planckian theory.  In Section \ref{setup} we implement the theory in a cosmological setting.  In Section \ref{standard} we discuss the  standard cosmological model.  In Section \ref{infesup} we explain how inflation provides a suppression mechanism. In Section \ref{darkenergy} we discuss a possible explanation for dark energy. In Section \ref{moreinflation} we discuss a possible mechanism to generate an early inflationary period. In Section \ref{conclusion} we present our conclusions.

\section{Trans-Planckian Theory}
\label{theory}

A field can be  expressed as a function of the space-time coordinates $\p(x^{\mu})$ or of the energy-momentum coordinates $\tp(p^{\mu})$.
The two formulations are related by the Fourier transform
\bea
\label{fourier}
\widetilde{\p}(p) &=&  M^2   \int \frac{d^4 x}{(2 \pi )^{2} } \ e^{ - i p^{\mu} x_{\mu}  } \ \p(x) \ , \nonumber \\
\p(x) &=& \frac{1}{M^2} \int \frac{d^4 p}{(2 \pi )^{2} } \ e^{ \ i p^{\mu} x_{\mu}  } \ \widetilde{\p}(p) \ ,
\eea
where $M$ is a normalization  constant.
This transformation  preserves the  quadratic mass term 
\beq
M^2 \int d^4 x \  |\p(x)|^2 = \frac{1}{M^2} \int d^4 p \ |\widetilde{\p}(p)|^2 
\ . \label{norms}
\eeq
We thus say that the mass term is self-dual.
An  ordinary  action is usually  an  integral over space-time of a certain Lagrangian density.  
For example, for a free  massless  scalar field,  the action is $S=  \int d^4 x \ \partial_{\mu}  \p^* \partial ^{\mu} \p $.  This expression, unlike the mass term,  is not self-dual under the Fourier transform.  
The dual version of $S$, which is $\tS=  \int d^4 p \ \partial_{\mu}  \tp^* \partial ^{\mu} \tp $,  is in fact a completely  different functional. 
The simplest  way to construct a  self-dual theory is to combine the two actions in a total action $\S$:
\beq
\label{sum}
\S =   S +   \widetilde{S} \ .
\eeq
This prescription, applied to the free massless scalar field, gives exactly the relativistic harmonic oscillator discussed before (\ref{rhoaction}).

The problem now is how to define a  ``generalized Fourier transform'' which is valid also in the presence of gravity and gauge interactions.
This  mathematical construction has to retain both diffeomorphism and gauge invariances. 
For this construction, we need to consider two distinct  manifolds, one for the coordinates  $x^{\mu}$ and one for the momenta $p^{\mu}$.
Both manifolds have  a  metric and  gauge connections.  Matter fields exist in both manifolds and the two expressions $\p(x)$ and $\tp(p)$ are related by the generalized version of the Fourier.  The scheme of interaction between the degrees of freedom is thus as follows:
\beq
\begin{array}{ccc}
 { matter} &  \Longleftrightarrow   & \widetilde{ matter} \\ [1.5mm]
 \updownarrow & & \updownarrow        \\[1.5mm]
 { gauge/gravity}  & \   & \ \, \widetilde{ gauge}/ \widetilde{ gravity}\ .\\
\end{array} 
\eeq
The case we consider in this paper is the simplest possible one, with just one scalar $\p$ as  matter field and one Abelian $U(1)$ gauge structure plus gravity:
\beq
\begin{array}{ccc}
 \p(x) &  \Longleftrightarrow   &  \tp(p)   \\ [1.5mm]
 \updownarrow & & \updownarrow        \\[1.5mm]
g_{\mu\nu}(x) , \, A_{\mu}(x) & \   &  \ \, \tg_{\mu\nu}(p),\,  \tA_{\mu}(p)  \ ,  \\
\end{array}
\eeq
$g_{\mu\nu}$ and $\tg_{\mu\nu}$ are the two metrics and $A_{\mu}$ and $\tA_{\mu}$ are two  U$(1)$  gauge bosons living on the two  manifolds $x$ and $p$. $g_{\mu\nu}$ and $\tg_{\mu\nu}$, as well as $A_{\mu}$ and $\tA_{\mu}$, are distinct degrees of freedom. They can influence each other only through the matter field.
The U$(1)$ gauge structure is rather special. It is not meant to be the electromagnetic gauge interaction, but instead something that couples universally to all matter fields with the same charge: a sort of ``graviphoton''. It will be needed to implement translational invariance in the cosmological solution.

$S$ is the usual local action in space-time variables that we are accustomed to:
\beq
\label{gravity}
S =  \int d^4 x \sqrt{-g} \ \left(  - \frac{1}{16 \pi \GN} R  -\frac{1}{4 e^2} F^2 +  \nabla_{\mu} \p^* \nabla^{\mu} \p   - m^2 |\p|^2 \right) \ ,
\eeq 
where the covariant derivative is  $\nabla_{\mu}= \frac{\partial}{\partial x^{\mu}} -i A_{\mu} $. 
$\GN$ is the Newton constant and the related Planck mass is $\MP = 1/ \GN^{1/2}$. The ratio $\MP/M $ remains a free parameter of the model.
On the trans-Planckian side, the action is the dual version of the previous one:
\beq
\label{gravitydual}
\widetilde{S} =  \int d^4 p \sqrt{-\widetilde{g}} \ \left( - \frac{1}{16 \pi \tGN } \widetilde{R}-\frac{1}{4 e^2} \tF^2  +  \nabla_{\mu} \widetilde{\p}^* \nabla^{\mu} \widetilde{\p}  -  \tm^2 |\tp|^2  \right) \ ,
\eeq
where the covariant derivative is  $\nabla_{\mu}=\frac{\partial}{\partial p^{\mu}} -i \tA_{\mu} $.
$\tGN$ is the dual Newton constant that defines a dual Planck mass $\tMP =1/ \tGN^{1/2}$. 
As a consequence of self-duality, we have that $\tGN=M^4 \GN$, $\tMP=\MP/M^2$ and $\tm = m/M^2$.

To define the generalised Fourier transform, we need to impose a  restriction on the metric. This can be done by choosing a suitable boundary condition. There must be certain space slice, which we denote as $*$,   for both the  manifolds $x^{\mu}$ or $p^{\mu}$,  where the  metrics $g _{\mu\nu}(0)$ and $ \tg _{\mu\nu}(0) $ s can be written as $\eta_{\mu\nu}$.

We then need to introduce two auxiliary flat Minkowski spaces, which we denote as $y_{\nu}$ and $q_{\nu}$.
The matter field can written as a function over these auxiliary spaces $\p(y)$ and $\widetilde{\p}(q)$, and the two are related by the ordinary Fourier transformation (\ref{fourier}).
To obtain the fields $\p(x)$ and $\widetilde{\p}(p)$, we use the following embedding formulae:
\bea
\p(x) &=&  \int \frac{d^4 q} {(2 \pi )^{2} } \ f_{q}(x) \widetilde{\p}(q)\ \ , \nonumber \\
\widetilde{\p}(p) &=&  \int \frac{d^4 y} {(2 \pi )^{2} } \ \tf_{y}(p) \p(y)\ \label{covfourier} \ ,
\eea
where we call $f_{q}(x)$ and $\tf_{y}(p)$  the {\it  probe functions}.
The problem of finding a covariant Fourier transform has now become that of finding a proper covariant definition for  $y_{\nu}$ and $q_{\nu}$ and for the probe functions.

\begin{figure}[h!t]
\epsfxsize=11.5cm
\centerline{\epsfbox{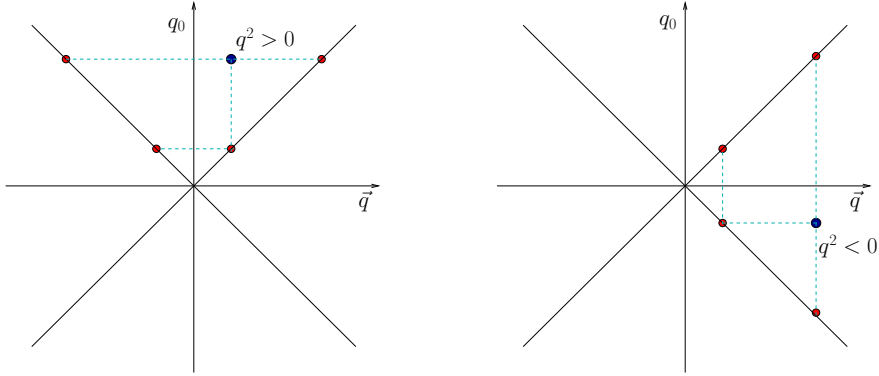}}
\caption{{\footnotesize  Light-cone decomposition for off-shell probe functions.}}
\label{othernew}
\end{figure}
Let us focus on the  probe function $f_{q}(x)$.
We can decompose the vector $q$ as the sum of null vectors.  
For time-like vectors $q^2>0$, we decompose the four-vector $(q_0,\vec{q})$ in the following sum of null vectors,  as described in the left side of Figure \ref{othernew} 
\beq
\label{decomposeqdue}
(q_0,\vec{q}) = \frac{1}{2}(q_0,  \hat{q} q_0) + \frac{1}{2}(q_0,-  \hat{q}q_0 ) + \frac{1}{2} (|\vec{q}|,\vec{q}) - \frac{1}{2}(|\vec{q}|,-\vec{q}) \ .
\eeq
The corresponding probe function can be defined as follows
\beq
\label{secondnewpresc}
f_{(q_0,\vec{q})}(x) = f_{(q_0, \hat{q} q_0) }(x)^{1/2} f_{(q_0,-  \hat{q}q_0 )}(x)^{1/2} f_{(|\vec{q}|,\vec{q})}(x)^{1/2}/f_{(|\vec{q}|,-\vec{q})}(x)^{1/2} \ .
\eeq
This definition is self-consistent: as $q$ approaches the light-cone, $f_{(q_0,-  \hat{q}q_0 )}(x)^{1/2}$  cancels with  $f_{(|\vec{q}|,-\vec{q})}(x)^{1/2}$, and what is left is precisely $f_q$. 
For space-like probes, we use instead the decomposition in the right side of Figure \ref{othernew}.

We finally define the light-like probe functions, the ones for which $q^{\mu}$ is a null vector. 
We write an auxiliary  action for the probe function $f_q(x)$ with $q^2=0$ as follows:
\beq
\label{actioneprobe}
{\cal S}_{f_q}  =   \int d^4 x \sqrt{-g} \ \left(   \nabla_{\mu} f_q^* \nabla^{\mu} f_q  -   \bar{A}^{\mu } (i f_q^* \nabla_{\mu} f_q  + { h.c.} ) + \bar{A}_{\mu}^2   f_q ^* f_q  \right) \ .
\eeq
The  covariant derivative $\nabla_{\mu}$ assures that the gauge invariance is satisfied. 
We have introduced another vector field $\bar{A}_{\mu}$ that is defined as follows.
$\bar{A}_{\mu}$ is a  vector field defined everywhere in space-time that coincides with $A_{\mu}$ on the spacial slice $*$, i.e., $\bar{A}_{\mu}(*)=A_{\mu}(*)$.
While $A_{\mu}$  transforms under gauge transformations, $\bar{A}_{\mu}$ does not. Moreover,  $\bar{A}_{\mu}$ is not a dynamical field; it  is just a canonical extension of the boundary condition $\bar{A}_{\mu}(*)$ to the whole of space-time. We define this canonical extension by keeping $\bar{A}_{\mu}$ constant along geodesics.
A detailed description of this will be given in the next section.
If $A_{\mu}$ coincides with   $\bar{A}_{\mu}$ in a certain region of space-time, and in a certain gauge, the equation for $f$  derived from the action (\ref{actioneprobe}) simply reduces to
\beq
\label{eomf}
g^{\mu \nu} \partial_{\mu} \partial_{\nu} f_q(x) + \frac{\partial_{\nu} (\sqrt{-g} g^{\mu \nu})}{\sqrt{-g}} \partial_{\mu} f_q(x)   =  0 \ .
\eeq

The generalised Fourier transform consists of a chain of relations connecting $\p(x)$ to $\tp(p)$ passing through $\tp(q)$ and $\p(y)$:
\beq
\begin{array}{ccc}
 \tp(q) &  \longleftrightarrow   &  \p(y)   \\ [1.5mm]
 \updownarrow & & \updownarrow        \\[1.5mm]
\p(x) & \   & \ \,  \tp(p) \ .  \\
\end{array} 
\eeq
The formulas to invert (\ref{covfourier}) can be formally written as
\bea
\widetilde{\p}(q)  &=&   \int \frac{d^4 x \sqrt{-g}} {(2 \pi )^{2} } \ f_{q}^{-1}(x) \p(x)\ \ , \nonumber \\
\p(y) &=&   \int \frac{d^4 p \sqrt{-\tg} } {(2 \pi )^{2} } \ \tf_{y}^{-1}(p) \widetilde{\p}(p) \label{inversionfourier} \ \ .
\eea
In  flat space, the inverse of a probe function $f^{-1}$ coincides with the complex conjugate $f^*$, but this is not true for a  general metric.  
The functions $f_q^{-1}(x)$ and $\tf_y^{-1}(p)$ are defined to satisfy the orthogonality relations:
\bea
\label{inv}
 \frac{1}{\sqrt{-g}} \delta(x-x')  &=&   \int \frac{d^4 q  } {(2 \pi )^{4} } \ f_{q}(x) f_{q}^{-1}(x') \ , \nonumber  \\
 \frac{1}{\sqrt{-\tg}} \delta(p-p')  &=&   \int \frac{d^4 y} {(2 \pi )^{4} } \ \tf_{y}(p) \tf_{y}^{-1}(p') \label{inversionfourierdeltafunction} \ .
\eea
Later, when  we will need the inverse probe functions $f^{-1}$, we will compute them  in an adiabatic limit in which the variation of the metric is not too rapidly with respect to the probe wavelength.

The total action $\S$ can be expressed uniquely as a function of $\p$ by using the inversion formulas above. When the action is quadratic in the matter field, we have 
\bea
\label{total}
\S = \int d^4 x \sqrt{-g}\ \left(  \nabla_{\mu} \p^* \nabla^{\mu} \p   - m^2 \p^*\p + \int d^4 x' \sqrt{-g'}\ \p^*(x'){\cal F}(x',x)  \p(x)  \right) \ ,
\eea 
where ${\cal F}$ is given by
\bea
{\cal F}(x',x) &=&  \int  \frac{d^4 q'} {(2 \pi M )^{2} }   \frac{ d^4 y'} {(2 \pi )^{2} }  \frac{d^4 p \sqrt{-\tg} } {(2 \pi )^{4} }   \frac{ d^4 y} {(2 \pi )^{2} }   \frac{d^4 q} {(2 \pi M )^{2} }   \nonumber \\ [1.5mm]
&&  \quad   \left( f_{q'}^{-1 *}(x') e^{iq'y'}  (\tnabla_{\mu} \tf_{y'}(p))^* 
 \tnabla^{\mu} \tf_y(p) \ 
e^{-iqy}\ 
 f^{-1}_q(x) \parlarge  \right. \nonumber \\ 
[1mm]   &&  \qquad \left. \parlarge - \tm^2  f_{q'}^{-1 *}(x') e^{iq'y'}    \tf_{y'}(p)^* 
  \tf_y(p) \ e^{-iqy}\ f^{-1}_q(x) \right) \ .
\label{inversione}
\eea

The field equations for the metric $g_{\mu \nu}$ are the usual Einstein equations, $G_{\mu\nu} = 8 \pi \GN T_{\mu \nu}$, where the tensor $T_{\mu \nu}$ is given by
\beq
\label{emtensor}
\int d^4x \frac{1}{2} \sqrt{-g} \  T_{\mu \nu} = \frac{\delta \S_{ mat}}{\delta g^{\mu\nu}} =  \frac{\delta S_{ mat}}{\delta g^{\mu\nu}} \ .
\eeq
The second passage in this formula follows from the fact that the dual-action $\widetilde{S}$ is, by definition, independent of the space-time metric $g_{\mu \nu}$. 
The energy-momentum tensor, neglecting the kinetic term for the gauge fields,  is
\beq
\label{energymomentumtensor}
T_{\mu \nu} = 2 (\nabla_{\mu}  \p)^* \nabla_{\nu} \p - g_{\mu\nu} \Big( g^{\alpha \beta} (\nabla_{\alpha}  \p)^* \nabla_{\beta} \p  - m^2 |\p|^2\Big) \ ,
\eeq
and is the ordinary energy-momentum tensor as computed from the $S$ part of the action.
The only way the dual-metric $\tg_{\mu \nu}$ can interact with the metric $g_{\mu \nu}$ is indirectly, by changing the equation of motion for the matter fields. 
Dual-gravity satisfies the dual-Einstein equations $\widetilde{G}_{\mu\nu} = 8\pi \tGN \widetilde{T}_{\mu\nu}$, and the tensor $\widetilde{T}_{\mu\nu}$ is obtained from $\delta \widetilde{S}/\widetilde{g}^{\mu\nu}$, similar to equation (\ref{emtensor}).

\section{Cosmology}
\label{setup}

We now implement the previous theory in a cosmological setting.
We consider first the case of a homogeneous and translational invariant universe  described by the FRW metric  
\beq
\label{FRW}
ds^2 =  dt^2 -  a(t)^2  d \vec{x}^2  \ ,
\eeq
where $t$ is the time coordinate, $a(t)$ the expansion factor, and $\vec{x}$ the comoving spatial coordinates.
We must also consider a  dual universe with  the same type of metric 
\beq
\label{FRWduale}
d\widetilde{s}^2 =  de^2 - {\ta}(e)^2   d \vec{p}^2 \ ,
\eeq
where $e$ is the energy, $\vec{p}$ the comoving three momentum, and ${\ta}(e)$ the dual-expansion factor.
We have taken zero spatial curvature in the FRW metrics. 
The universes are asymptotically flat if
\beq
\label{asymptoticflatness}
\lim_{t \to \infty} \frac{\dot{a}(t)}{a(t)} = 0 \ ,  \qquad \qquad \lim_{e \to \infty} \frac{\dot{\ta(e)}}{\ta(e)} = 0 \ .
\eeq

The boundary conditions needed to implement the generalized Fourier transform are set at $t=0$ and $e=0$.
For the gauge fields, we choose the following conditions in those particular spatial slices
\bea
\label{ansatzgauge}
 &&A_{\mu}(0,\vec{x}) = (0,\vec{x})\ , \qquad \qquad  \tA_{\mu}(0,\vec{p}) =  (0,\vec{p}) \ , \nonumber \\
  &&F_{\mu\nu}(0,\vec{x}) = 0\ , \qquad \qquad \quad \ \  {\tF}_{\mu\nu}(0,\vec{p}) =  0 \ .
\eea
We declare $\bar{A}_{\mu}$ and $\bar{\tA}_{\mu}$ to be 
\beq
\label{ansatzgauged}
 \bar{A}_{\mu}(t,\vec{x}) = (0,\vec{x})\ , \qquad \quad \qquad \bar{\tA}_{\mu}(e,\vec{p}) =   (0,\vec{p}) \ .
\eeq
$A_{\mu}$ and $\tA_{\mu}$ are dynamical fields but for the rest of the paper, we will make the assumption that $A_{\mu} =  \bar{A}_{\mu} $ and $\tA_{\mu}=\bar{\tA}_{\mu}$ everywhere in space-time. This simplifies the task of computing the probe function.

We will do the analysis for the probe functions $f_q(x)$ explicitly, the same will apply to the dual probes.  We have to define the  light-like probes with  $q_{\mu}=(\omega = |\vec{q}|, \vec{q})$; the others follows from the decomposition in null vectors given earlier.
The spatial slice $t = 0 $ is where the momentum $q_{\mu}$ is defined.
The equation we have to solve is (\ref{eomf}). Since we have translational invariance, we can separate the variables and use the following ansatz:
\beq
\label{probecosmo}
f_q(x) = e^{- i \vec{q}\vec{x} } f_{\omega}(t)\ ,
\eeq
where the time-dependent probe function $f_{\omega}(t)$ satisfies the following equation:
\beq
\label{reducedprobecosmo}
\partial_t(a(t)^3 \partial_t f_{\omega}(t)) + \omega^2 a(t) f_{\omega}(t)  = 0 \ ,
\eeq
with the boundary condition 
\beq
f'_{\omega}(0) = \frac{i \omega}{a(0)} f_{\omega}(0)\ 
\eeq
with $f_{\omega}(0)$ real.
The last is the conditions that the probe functions are all synchronized at $t=0$ for every $\omega$. 
The overall normalization of the probe function can be chosen to be arbitrary.

If the asymptotic flatness  condition is satisfied (\ref{asymptoticflatness}), the  late-time  solution  is in general in the following form:
\beq
\label{fadiabatic}
f_{\omega}(t) \simeq \frac{1}{a(t)} e^{i \delta} e^{  i  \omega t / \gamma a(t)  }  + r \frac{1}{a(t)}  e^{- i \delta} e^{  - i  \omega t / \gamma a(t)  } \ , \quad  \qquad  t \to \infty \ .
\eeq
The factor $a$ in the denominator of the exponent is a red-shift factor caused by the cosmological expansion, and $\gamma$ is a constant that depends on the specific function $a(t)$. The modulus of the probe function $h(t)$ has, in general, a non-trivial time dependence. 
In flat space, the probe function would be simply $f_{\omega}(t) \simeq   e^{  i  \omega t   }$. 
In general, we also expect  a scattering phase $\delta$ and the scattering  component $ \propto r   e^{ - i  \omega t }$   due to the interaction with the metric.

For generic power-law behaviour of the scale factor $a(t) \propto t^{\alpha}$, the solution for the probe function can be explicitly written  in terms of the Bessel function:
\beq
f_{\omega}(t) \propto  t^{(1-3\alpha)/2} J_{\pm\frac{3\alpha-1}{2(\alpha-1)}}\left(\frac{\omega t^{1-\alpha}}{1-\alpha}\right)
\eeq
and at large $t$, they behave as
\beq
\label{genericalphaasymt}
f_{\omega}(t) \simeq  \frac{1}{t^{\alpha}} e^{\pm  i t^{1-\alpha} \omega/(1- \alpha)} \ .
\eeq
Important cases for later use are   $\alpha=1/2$ and $\alpha=2/3$, which correspond to the  radiation and  matter-dominated universes. 
In both cases, the scattering phase does not depend on $\omega$ and the scattering back component is zero $r=0$.

To write  dual-action $\tS$ as a function of $\p(x)$, we need to invert the probe functions. 
We can do this explicitly in the adiabatic limit.
From the  orthogonality relation  
\beq
\int dt a(t)^3 \ h(t)e^{i t \omega/a(t)} \  \frac{e^{-i t \omega'/a(t)}}{a(t)^4 h(t)} = 2\pi  \delta(\omega-\omega') \ ,
\eeq
 we have
\beq
f_{\omega}(t) =  h(t)e^{i t \omega/a(t)} \ , \qquad \qquad  f_{\omega}^{-1}(t) =  \frac{e^{-i t \omega/a(t)}}{a(t)^4 h(t)}  \ .
\eeq 
For the dual probe function $\tf$, we have:
\beq
\tf_{\tau}(e) =  \th(e)e^{-i e \tau/\ta(e)} \ , \qquad \qquad  \tf_{\tau}^{-1}(e) =  \frac{e^{i e \tau/\ta(e)}}{\ta(e)^4 \th(e)}  \ .
\eeq

\begin{figure}[h!t]
\epsfxsize=15cm
\centerline{\epsfbox{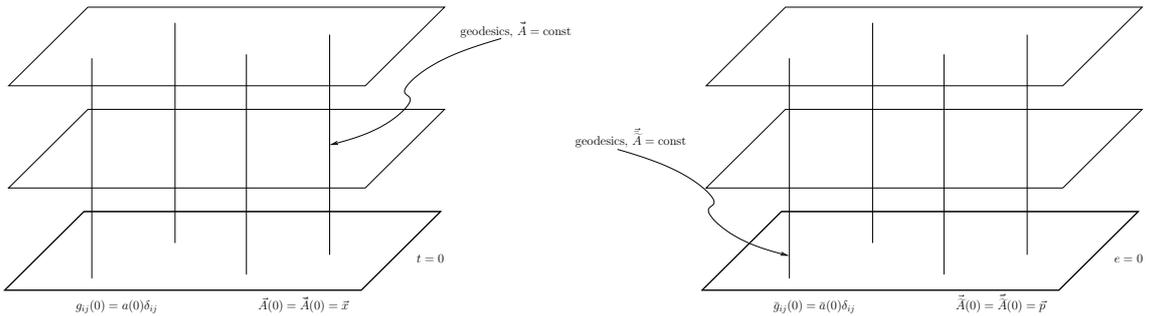}}
\caption{{\footnotesize General metrics and boundary conditions.}}
\label{general}
\end{figure}

The formalism can be applied to any cosmological metric (see Figure \ref{general}), not only the homogeneous FRW. 
We can always bring any metrics in the synchronous form
\beq
ds^2 =  dt^2 -  g_{ij}  dx^i dx^j  \ , \qquad \quad 
d\widetilde{s}^2 =  de^2 - {\tg}_{ij}   dp^i dp^j  \ .
\eeq
The boundary conditions we need are
\beq
 g_{ij}(0,\vec{x})=a(0)\delta_{ij} \ , \quad \qquad \tg_{ij}(0,\vec{p}) = {\ta}(0) \delta_{ij}\ ,
\eeq
together with (\ref{ansatzgauge}) and (\ref{ansatzgauged}). 

\section{Standard Cosmological Model}
\label{standard}

In ordinary gravity, multiplying the universe scale factor $a(t)$ by a constant $a(t) \to \lambda a(t)$ does not change the physics since it  can be compensated for  by  a coordinate redefinition.  In our model we have two scale factors $a(t)$ and $\ta(e)$. 
We can use the freedom of rescaling to fix
\beq
\label{choice}
M^2 a(t) = \ta(e) \ .
\eeq
After that, the overall scale of $a(t)$ becomes a physical observable and will enter into the final result.

In the standard cosmological model,  we have an initial stage which is radiation dominated followed by a second stage that is matter-dominated:
\bea
\label{scalefactor}
&&a(t)= c_{1/2} t^{1/2}  \qquad  \qquad \qquad \qquad t \leq t_{\rm eq} \nonumber \ , \\
&&a(t) = c_{2/3} \left(t + \frac{ t_{\rm eq}}{3}\right)^{2/3}   \qquad  \qquad  t \geq t_{\rm eq} \ ,
\eea
The transition between the two  happens at time $t_{\rm eq}$ when there is equivalence  between the matter and the radiation components.
The particle $\p$ that we considered in the previous actions is now identified with a cold  dark matter particle.
It  freezes-out at the time $t_m$, which is when the temperature of the universe reaches the scale $m$ of the particle mass. This time is approximately $t_m \simeq  \MP/m^2 $. 
Since $\p$ is not the only component of the universe, the equivalence time $t_{\rm eq}$ is in general different from the freeze-out time $t_m$. 
We denote as $10^{2 X}$ the  shift from $t_m$ to $t_{\rm eq}$ so that $a(t_{\rm eq})/a(t_m)=10^X$. $t_{\rm eq}$ and $c_{2/3}$ are then given by:
\beq
t_{\rm eq} = 10^{2X }t_{m} \simeq \frac{10^{2X} \MP}{m^2} \ , \qquad \qquad c_{2/3}  \simeq c_{1/2} \frac{m^{1/3}}{10^{X/3} \MP^{1/6}} \ .
\eeq
The dual universe has a specular behaviour
\bea
&&\ta(e)= \tc_{1/2} e^{1/2}  \qquad  \qquad \qquad \qquad e \leq e_{\rm  eq} \nonumber \ , \\
&&\ta(e) =  \tc_{2/3} \left(e + \frac{ e_{\rm eq}}{3}\right)^{2/3}   \qquad  \qquad   e \geq e_{\rm  eq} \ ,
\eea
with transition that happens at $e_{\rm  eq} \simeq 10^{2X} e_m$.
The choice (\ref{choice}) allows us to relate the various parameters as follows:
\beq
\tc_{1/2} = c_{1/2} M \ ,
\eeq
and also
\beq
e_{\rm  eq} = 10^{2X} e_{m} \simeq \frac{10^{2X} M^2 \MP}{m^2}  \ , \qquad \quad \tc_{2/3}  \simeq \tc_{1/2} \frac{m^{1/3}}{10^{X/3} M^{1/3} \MP^{1/6}} \ .
\eeq

The modulus of the probe functions, $h(t)$ in (\ref{fadiabatic}), has the following behaviour in the radiation and  matter-dominated universe:
\bea
\label{probeschangingbehaviour}
 && h(t) \simeq \frac{1}{t^{1/2}} \frac{m^{1/3} M^{1/3}}{10^{X/3} \MP^{1/6}}  \  \quad \qquad  \qquad\quad   t\leq t_{\rm eq} \nonumber \ , \\
&&  h(t)=\frac{M^{1/3}}{(t+ t_{\rm eq}/3)^{2/3}}    \qquad \qquad \qquad \quad  t\geq t_{\rm eq} \ ,
\eea
where  we have chosen the  normalization at our convenience.
The dual version of (\ref{probeschangingbehaviour}) is given by:
\bea
&&\th(e) \simeq \frac{1}{e^{1/2}} \frac{m^{1/3}}{10^{X/3} M^{2/3}  \MP^{1/6}} \  \qquad  \quad  \quad    e\leq e_{\rm  eq} \nonumber \ ,\\
&&\th(e) = \frac{1}{(e + e_{\rm eq}/3)^{2/3} M^{1/3}}      \qquad  \qquad  \quad   e \geq e_{\rm  eq}  \ .
\eea

We want to understand how the dual part of the action affects the equation of motion for the particle $\p$. For this, we need to compute explicitly the expression  (\ref{inversione}). 
We first do the computation  for the $0$-$0$ part of the kinetic term. 
This term is  $\int de \ta(e)^3 \partial_e \tp^* \partial_e \tp$ and then reduces to  the following chain of integrals:
\bea
\label{firstinversion}
\int de \ta(e)^3 \int d\tau' \int d\omega' \int dt' a(t')^3 \int  d\tau \int d\omega \int dt a(t)^3 \frac{1}{(2\pi)^3} \p^*(t') \p(t) \nonumber \\ 
\frac{\tau \tau'}{\ta(e)^2} \ \ \ \th(e) e^{i \tau' e /\ta(e)} e^{i\omega' \tau'} \frac{e^{i t' \omega'/a(t')}}{h(t') a(t')^4} \ \ \ \th(e) e^{-i \tau e /\ta(e)} e^{-i\omega \tau} \frac{e^{-i t \omega/a(t)}}{ h(t)a(t)^4} \ . \ \ 
\eea
We do first the integral $de$  by  isolating the terms that depend explicitly on $e$
\beq
\label{eintegral}
\int de \  \ta(e)^3  \frac{1}{\ta(e)^2} \th(e) e^{i \tau' e /\ta(e)} \th(e) e^{-i \tau e /\ta(e)}   \ .
\eeq
This can be expressed as 
\beq
\label{approximationdelta}
\int  ds \frac{c_{1/2}^2 M^{2/3}  m^{2/3}}{ 10^{2X/3}  \MP^{1/3}} e^{i s (\tau'-\tau)} =  \frac{c_{1/2}^2 M^{2/3} m^{2/3}}{ 10^{2X/3} \MP^{1/3}} 2 \pi \delta(\tau-\tau') \ ,
\eeq
where we changed variable from $e$ to $s = e/\ta(e)$.
The next step  is to  integrate $d\tau'$, and the main integral (\ref{firstinversion}) reduces to the following:
\bea
 \int d\omega' \int dt' a(t')^3 \int d\tau \int d\omega \int dt a(t)^3 \frac{1}{(2\pi)^2}   \p^*(t') \p(t) \nonumber \\  \frac{c_{1/2}^2 M^{2/3} m^{2/3}}{ 10^{2X/3} \MP^{1/3}}  \tau^2 \   e^{i\omega' \tau} \frac{e^{i t' \omega'/a(t')} }{h(t')a(t')^4} \ \  e^{-i\omega \tau} \frac{e^{-i t \omega/a(t)} }{h(t) a(t)^4} \label{secondstep} \ . \ \ 
\eea
Then we integrate $d \tau$, whose only dependent part is
\beq
\int d\tau \tau^2 e^{i \tau (\omega'-\omega)} = - 2\pi \delta''(\omega - \omega') \ ,
\eeq
where $\delta''$ is the second derivative of the delta function. Then we integrate $ d\omega'$, and (\ref{secondstep}) reduces to
\bea
 \int dt' a(t')^3 \int d\omega \int dt a(t)^3  \frac{1}{2\pi}   \p^*(t') \p(t) \ \ \  \nonumber \\  \frac{c_{1/2}^2 M^{2/3} m^{2/3}}{ 10^{2X/3} \MP^{1/3}}  \ \  \frac{t'^{2}}{a(t')^2}  \frac{e^{i t' \omega/a(t')}}{h(t') a(t')^4} \ \ \  \frac{e^{-i t \omega/a(t)}}{h(t) a(t)^4} \ . \label{thirdstep}
\eea
Then is the turn of the  $\omega$ dependent part which gives 
\beq
\int d \omega e^{i \omega (t/a(t) -t'/a(t'))} =2\pi a(t) \delta(t-t') \ .
\eeq
Finally, we integrate $dt'$ so that (\ref{thirdstep}) becomes
\beq
\label{resultzerozeroterm}
 \int dt a(t)^3 \    \frac{c_{1/2}^2 M^{2/3} m^{2/3}}{ 10^{2X/3} \MP^{1/3}}   \  \frac{t^{2}}{ h(t)^2 a(t)^6} \  \p^*(t) \p(t) \ .
\eeq
This completes the inversion of the original expression  (\ref{firstinversion}).  This can also be rewritten, by using (\ref{scalefactor}) and (\ref{probeschangingbehaviour}), in a more convenient form  which in the radiation-dominated period is
\beq
\label{negativemasssquareterm}
 \int dt a(t)^3  \\  \frac{1}{c_{1/2}^4}  \p^*(t) \p(t) \ ,
\eeq
and in the matter-dominated period is
\beq
\label{negativemasssquaretermmatterlater}
 \int dt a(t)^3  \\  \frac{10^{4X/3} \MP^{2/3}}{c_{1/2}^4 m^{4/3}(t+ t_{\rm eq}/3)^{2/3}}  \p^*(t) \p(t) \ .
\eeq
Note that this term counts as  a  {\it negative} mass squared term in the full action $\S$. The sign is positive since we started from a $0$-$0$ kinetic term in momentum space that had positive sign; the inversion makes it  a negative contribution to the potential energy. In flat space-time, as in  the relativistic harmonic oscillator (\ref{rhoaction}), this term was a negative potential, but proportional to $t^2$ and not constant.

Let us now invert also the other terms in $\tS$.  
The space part $i$-$i$ of the kinetic term is:
\bea
 -\int de \ta(e)^3 \int d\tau' \int d\omega' \int dt' a(t')^3 \int d\tau \int d\omega \int dt a(t)^3 \frac{1}{(2\pi)^3}   \vec{\partial}\phi(t')^* \vec{\partial}\phi(t)  \nonumber \\  \frac{1}{\ta(e)^2}   \  \th(e) e^{i \tau' e /\ta(e)} e^{i\omega' \tau'} \frac{e^{i t' \omega'/a(t')}}{h(t') a(t')^4} \   \th(e) e^{-i \tau e /\ta(e)} e^{-i\omega \tau} \frac{e^{-i t \omega/a(t)}}{ h(t) a(t)^4} \ \ \ \ \ \ ,
\eea
where $\p = e^{i M^2 \vec{x}^2 /2}$.
The computation of the integrals is similar to the one before, and we just give the final result
\beq
 -\int dt a(t)  \\   \vec{\partial}\phi(t)^* \vec{\partial}\phi(t) \ .
\label{spacekinetikterminverted}
\eeq
This is valid both in the radiation-dominated and matter-dominated regions.
The dual-mass term which is given by 
\bea
-\int de \ta(e)^3 \int d\tau' \int d\omega' \int dt' a(t')^3 \int d\tau \int d\omega \int dt a(t)^3 \ \frac{1}{(2\pi)^3} \p(t')^* \p(t) \nonumber \\  \frac{m^2}{M^4} \  \th(e) e^{i \tau' e /\ta(e)} e^{i\omega' \tau'} \frac{e^{i t' \omega'/a(t')}}{h(t') a(t')^4} \ \ \ \th(e) e^{-i \tau e /\ta(e)} e^{-i\omega \tau} \frac{e^{-i t \omega/a(t)}}{h(t) a(t)^4} \ \ \ \ \  .
\label{firstpassagedualmass}
\eea
and the computing the integrals give 
\beq
\label{dualmassinvertedfinal}
 -\int dt a(t)^3  \\ c_{1/2}^4  m^2    \    \p(t)^*  \partial_t^2 \p(t) \ .
\eeq
More details about these computations can be found in \cite{Bolognesi:2012fq}.

 The first goal we want to achieve is to suppress $\tS$ with respect to $S$ at low energy. We see from the results above that is is not possible in a standard cosmological model. For example if we take a  very large coefficient $c_{1/2}$,  this would suppress the term  (\ref{negativemasssquaretermmatterlater}) but then (\ref{dualmassinvertedfinal}) would be a dominant term. Something more is needed to make $\tS$ invisible.

\section{Inflation and Suppression}
\label{infesup}

\begin{figure}[h!t]
\epsfxsize=8cm
\centerline{\epsfbox{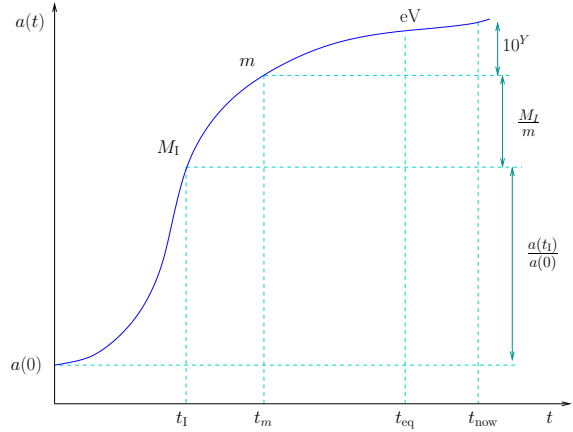}}
\caption{{\footnotesize Universe expansion with an inflationary stage.}}
\label{inflation}
\end{figure}

Inflation provides a viable suppression mechanism.
The various  stages of the universe expansion are summarized in Figure \ref{inflation}.
The expansion factor has first inflationary  stage  up to $t_{\rm I}$ and then enter into the standard cosmological expansion:
\beq
 a(t) =\left\{ \begin{array}{cc} 
 a(0) \  e^{ \sigma  t} \qquad &   t \leq t_{\rm I} \\ 
 c_{1/2} (t - t^*)^{1/2} \qquad &  t_{\rm I} \leq t \leq t_{\rm eq}\end{array} \right.
\eeq
For the moment we do  not address the mechanism that  generates  the inflationary stage, we just assume its existence; we will return to this point in Section \ref{moreinflation}.
$a(0)$, $\sigma$ and $t_{\rm I}$ are arbitrary parameters for the moment.
The total expansion factor during inflation is given by: 
\beq
\frac{a(t_{\rm I})}{a(0)} =  e^{\sigma t_{\rm I}}  \ .
\eeq
If we call  $M_{\rm I}$ the inflation energy scale, we have
\beq
\label{sigma}
\sigma \simeq \frac{M_{\rm I}^2}{\MP} \simeq \frac{1}{(t_{\rm I} - t^*)^2} \ .
\eeq
Usually inflation is considered to occur at the GUT scale at $10^{16}$ GeV, and we will also consider this value as a reference in what follows although the main result will not depend on this assumption. 
The other matching condition gives
\beq
 a(0) \  e^{ \sigma t_{\rm I} }  =  c_{1/2} \frac{\MP^{1/4}}{M_{\rm I}^{1/2}} \ .
\eeq

The solution of the  probe equation (\ref{reducedprobecosmo}), restricted to the inflationary stage, can be obtained analytically:
\beq
f_{\omega}(t) = e^{-\sigma t} \cos{\left(\frac{\omega(1-e^{-\sigma t})}{a(0) \sigma}\right)} + \frac{a(0) \sigma}{\omega} \sin{\left(\frac{\omega(1-e^{-\sigma t})}{a(0) \sigma}\right)} \ .
\eeq
We can consider the solution in two different regimes.
In the first, for $t \ll 1/\sigma$, there is an oscillatory behaviour
\beq
f_{\omega}(t) \simeq  \cos{\left(\frac{\omega  t}{a(0)}\right)}  + {\cal O}(\sigma t) \ ,
\eeq
and in the second, for  $t \gg 1/\sigma$, the solution saturates to a constant value
\beq
\label{constantapproach}
f_{\omega}(t) \simeq  \frac{a(0) \sigma}{\omega} \sin{\left(\frac{\omega}{a(0) \sigma}\right)} + {\cal O}(e^{-\sigma t}) \ .
\eeq
So $\p$ is frozen for sufficiently large time so that
\beq
\frac{a(0) \sigma}{\omega}  e^{\sigma t} \gg 1 \ .
\eeq
This is a very well-known effect in the theory of cosmological perturbations. 

Also the dual universe begin with  an inflationary phase from  $e=0$ to  $e_{\rm I}$
\beq
 \ta(e) =\left\{ \begin{array}{cc} 
 \ta(0) \  e^{ \ts  e} \qquad &   e \leq e_{\rm I} \\ 
 \tc_{1/2} (e - e^*)^{1/2} \qquad &  e_{\rm I} \leq e \leq e_{\rm eq}\end{array} \right.
\eeq
The solution for the dual probe function is
\beq
\tf_{\tau}(e) = e^{-\ts e} \cos{\left(\frac{\tau(1-e^{-\ts e})}{\ta(0) \tau}\right)} + \frac{\ta(0) \tau}{\ts} \sin{\left(\frac{\tau(1-e^{-\ts e})}{\ta(0) \ts}\right)} \ .
\eeq
and it saturates to a constant value for
\beq
\frac{\ta(0) \ts}{\tau}  e^{\ts e} \gg 1 \ .
\eeq

\begin{figure}[h!t]
\epsfxsize=14.5cm
\centerline{\epsfbox{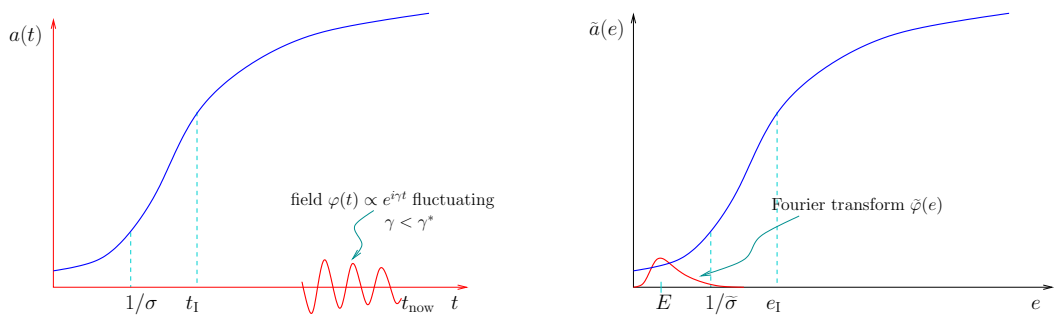}}
\caption{{\footnotesize The suppression mechanism works when the Fourier transform $\tp$ is contained in the dual early-inflationary stage.}}
\label{suppressioninflation}
\end{figure}

Now let us consider the wave  function in space-time $\p(t)$ and  assume that the equation of motion is dominated by $S$.     $\tS$ is  negligible for some reason to be uncovered.
The field is oscillating as $\p \propto e^{i  \gamma t }$ with some frequency  $\gamma$.
The frequency is given by  $\gamma \simeq \sqrt{\MP/(t-t^*)}$  for $t \leq t_m$ and  is decreasing like $1/a(t)$.  After $t \simeq t_m$  the matter field $\p$ oscillates with frequency fixed and equal to its mass $\p \sim e^{i m t}$.  
The generalized Fourier transform $\tp(e)$ is a spectral distribution peaking around a certain energy scale $e \simeq E$.  
To find $E$ as a function of $\gamma$ we have to use the chain of relations of  the generalised  Fourier transform.  
The probe functions in $t$ are  $ e^{i \omega t/a(t)}$, so $\omega = \gamma a(t)$.   $\omega$ is related to $\tau$ by the ordinary Fourier transform with $e^{i \omega \tau}$. Finally the dual probe function, assuming we are in an adiabatic limit, is  $e^{i \tau e/ \ta(e)}$.  Finally we have the relation between $E$ and $\gamma$:
\beq
\frac{E}{\ta(E) }  \simeq \gamma \,    a(t) \ .
\eeq
A significant difference occurs if $\gamma$ is bigger or smaller than $\gamma^*$ defined by
\beq
\label{gammastar}
\gamma^* =  \frac{1}{\ts \ta(0) \,  a(t)} \ .
\eeq
If $\gamma > \gamma^*$ the spectral distribution  $\tp(e)$ is outside the inflationary stage. In this case we can use the same computations we have done in the previous Section \ref{standard} for the standard cosmological model.
If  $\gamma < \gamma^*$ instead, the field  $\tp(e)$ is contained in a small region between $0$ and $1/ \ts$, which is the early-inflationary stage.  This last  case is the one described in  Figure \ref{suppressioninflation}.

A suppression mechanism  is at work if the distribution $\p$ is confined to   $e \leq 1/\ts$.
This is because the probes, in the adiabatic approximation, behave like $1/\ta(e)$ but during the inflationary stage they have their modulus frozen.
For the quadratic terms in the action, we lose  a factor of $\ta^2$  from the fact that $\tp$ is not growing because it is at the  super-horizon scale, and thus extremely non-adiabatic (\ref{constantapproach}).
The result is that $\tS$ passing from the after-inflation to the early-inflation zone is experiencing a total suppression which is quadratic with the scale factor:
\beq
\label{amountofsuppression}
\frac{\tS_{\gamma > \gamma^*}}{\tS_{\gamma < \gamma^*}} \simeq \left(\frac{\ta(e_{\rm  I})}{\ta(0)}\right)^2  \ .
\eeq
We can thus make  $\tS$ is negligible with respect to $S$, and thus unobservable, if the spectral distribution $\tp$ is concentrated in the early-inflationary stage.

\section{Dark-Energy}
\label{darkenergy}

To obtain the matter equation of state, we need  its  energy-momentum tensor evaluated on the solution of the equation of motion.      The  extra mass squared that we found in (\ref{resultzerozeroterm}), comes from the dual part of the action $\tS$ and thus  it does not affect directly  the energy-momentum tensor  (see Eqs.\ (\ref{emtensor}) and (\ref{energymomentumtensor})). 
It alters though the equation of motion of the matter field and so  its equation of state indirectly.  We thus use a the trick to  add and subtract  this extra mass squared term to the energy-momentum  tensor, thus rewriting (\ref{energymomentumtensor}) as follows:
\bea
\label{energymomentumtensoraddandsubtract}
T_{\mu \nu} & = & 2 (\nabla_{\mu}  \p)^* \nabla_{\nu} \p - g_{\mu\nu} \Big(g^{\alpha\beta} (\nabla_{\alpha}  \p)^* \nabla_{\beta} \p  - m_{ eff}^2 \p^* \p \Big) \nonumber \\ && + \,  g_{\mu\nu}  \, \delta m^2 \p^*  \p  \ ,
\eea
where the effective mass is  $m_{\rm eff}^2 = m^2 - \delta m^2$, and the difference $\delta m^2 $  comes from the extra mass term coming from $\tS$.
The first line of (\ref{energymomentumtensoraddandsubtract}) is that of a massive particle with mass $m_{\rm eff}$, whereas the second line is the same of  a  positive cosmological constant term  being  proportional to $g_{\mu\nu}$. 
The ratio of the cosmological constant and dark matter component in the universe now is roughly $\Omega_{\Lambda}:\Omega_{m} \simeq  0.7 : 0.2$, and, thus,  of the same order of magnitude. 
We thus want $\delta m^2$ to be of the same order of magnitude of  the mass squared of the dark matter particle.  We  rewrite (\ref{negativemasssquaretermmatterlater}) as
\beq
\label{negativemasssquaretermmatterconvenient}
 \int dt a(t)^3  \\  \frac{1}{c_{1/2}^4 }  \frac{a(t_{\rm eq})}{a(t_{\rm now})}  \p^*(t) \p(t) \ .
\eeq
If we want this negative mass squared term to be comparable in absolute value with $m^2$  we need to impose the following condition on $c_{1/2}$:
\beq
\label{conditionc}
c_{1/2} \simeq   \left(\frac{a(t_{\rm eq})}{a(t_{\rm now})}\right)^{1/4}  \frac{1}{ m^{1/2}} \simeq \frac{1}{10 \  m^{1/2}} \ .
\eeq

 $\gamma^*$ must become of the order of the dark matter mass $m$ exactly at the present cosmological epoch $t_{\rm now}$. This means that $\tp$ comes out of the suppression zone, and the effect of $\tS$ becomes observable, just now. 
 The magnitude of $\tS$, when $\tp$ comes out of the suppression zone, must justify the observed dark energy value (\ref{conditionc}). For the first two requirements to be compatible,  we also  have to assume that the dark matter mass $m$ is bigger than any energy scale observed so far.  For example, a WIMP dark matter can have a mass up to $300$ TeV  so there is plenty of possibilities to satisfy these requirements. 
\begin{figure}[h!t]
\epsfxsize=11cm
\centerline{\epsfbox{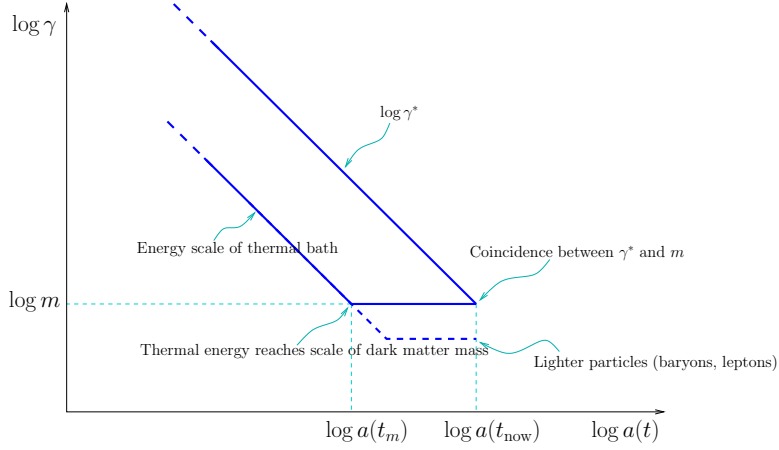}}
\caption{{\footnotesize The scale $\gamma^*$ and the scale of dark matter compared. Below $m_{max}$ line, the spectral distribution is included in the pre-inflation zone. Dark matter $\p$ becomes flat after the temperature reaches its mass scale. We want the intersection to coincide with the present cosmological epoch.   }}
\label{scalescoincidence}
\end{figure}
This mechanism is shown in Figure \ref{scalescoincidence}, where the upper bound is given by the line $\gamma^*$, decaying like $1/a(t)$, that crosses the dark matter scale at a particular time which we want to impose to be roughly  $t_{\rm now}$.  Lighter particles are still below the $\gamma^*$ bound and so do not yet feel the effect of $\tS$.

Using $\ta  = a  M^2$ and $\sigma = M^2 \ts$, the formula for the upper bound (\ref{gammastar}) is:
\beq
\label{mlightnew}
\gamma^* \simeq \frac{1}{\sigma  a(0) \,  a(t)} \ .
\eeq 
Now  we require $\gamma^* \simeq  m$ at $t_{\rm now}$, the coincidence of Figure \ref{scalescoincidence}, 
\beq
m \simeq  \frac{1}{\sigma a(0) \,  a(t_{\rm now})} \ ,
\eeq
and we have that
 \beq
\label{conditionsdualbecomesrelevant}
a(0) c_{1/2} \simeq \frac{1}{\sigma \MP^{1/2} 10^{Y}} \ .
\eeq
 $10^Y$ is the total expansion between the dark matter scale $t_m$ and now ($Y \simeq 16$ taking $m \simeq $ TeV). We have divided the total expansion  to factorise  $m$  from the equation.
Combining with the  requirement  $c_{1/2} \simeq  a(t_{\rm eq})^{1/4}/ a(t_{\rm now})^{1/4} m^{1/2} $ from the condition (\ref{conditionc}), we thus have 
\beq
\label{condition}
\frac{a(t_{\rm I})}{a(0)} \simeq\frac{ a(t_{\rm eq})^{1/2} M_{\rm I}  10^{Y} }{ a(t_{\rm now})^{1/2} m }    \ .
\eeq

We take  $M_{\rm I} \simeq M_{\rm GUT} \simeq M \simeq 10^{16}$ GeV. The mass $m$ of the cold dark matter is taken to be at the  TeV scale.  It  is important to have it bigger than the observable energy scales because ordinary massive particle should be still well inside  the suppression zone, with  only dark matter  coming out of the suppression zone.  $Y \simeq 16$ is the ratio of the TeV scale and the radiation temperature now.  These numbers  give 
\beq
\frac{a(t_{\rm I})}{a(0)}
\simeq 10^{27} \ ,
\eeq 
which is compatible with the number of e-folds required to solve the horizon problem.

In general, to solve the horizon problem,  we want the number of e-folds during inflation to be  equal to or  greater than the number of e-folds after inflation (with a small correction for the  matter-domination period which count as half of the other).  So the  correct  number of e-folds necessary to include the present horizon in the causal region is $a(t_{\rm eq})/a(t_{\rm I})$ for the radiation-dominated period multiplied by $(a(t_{{ now}})/a(t_{\rm eq}))^{1/2}$ for the matter-dominated period. 
Now we can rewrite the total expansion factor which is required to solve the horizon problem as:
\beq
\label{minimalforinflation}
\frac{a(t_{\rm I})}{a(0)} \simeq \frac{a(t_{\rm eq})^{1/2} a(t_{\rm now})^{1/2}}{a(t_{\rm I})}  \simeq  \frac{ a(t_{\rm eq})^{1/2}M_{\rm I} 10^{Y} }{ a(t_{\rm now})^{1/2} m}  \ .
\eeq
This is precisely equal to  (\ref{condition}) which was independently obtained from the conditions to solve for the dark energy problem.   The equivalence of the two conditions  does not depend on the particular choice of $M_I$, $M$, $\MP$ and $m$.

We expect that something new will happen at extremely low energies. 
The suppression mechanism caused by inflation gives
\beq
\frac{\tS}{S} \simeq \frac{ a(t_{\gamma})^2 }{ a(t_{\rm  I})^2} \frac{m^2}{\gamma^2} =\frac{ a(0)^2 }{ a(t_{\rm  I})^2 } \frac{m^4}{\gamma^4} \simeq 10^{-6} \frac{{\rm eV}^4}{\gamma^4} \ ,
\eeq
where in the last passage we used  $m  \simeq $ TeV and  GUT scale inflation.
For all the ordinary matter particles $\tS/S$ is negligible.

\section{More on Inflation}
\label{moreinflation}

Now we study the behaviour matter fields during  the inflationary stage. 
From $t_{\rm I}$ to $t_{\rm now}$ the equation of motion for $\p$ is dominated by $S$ and not $\tS$.
In particular the matter field at  $t_{\rm now}$ is approximately 
\beq
\p(t) \simeq  \frac{\MP}{m t} e^{i m t} \ .
\eeq
The generalized Fourier transform gives the dual field 
\beq
\label{tpdual2}
\tp(e) = \frac{1}{(2 \pi)^{3/2}}  \int d \tau \tf_{\tau}(e) \int d\omega e^{i \omega \tau} \int dt a(t)^3 f_{\omega}^{-1}(t)\p(t) \ ,
\eeq
This field had support in the dual inflationary stage (see Figure \ref{suppressioninflation}). We thus have 
\beq
\tf_{\tau}(e) \simeq \frac{1}{e_{\rm I}^{1/2}} \frac{m^{1/3}}{10^{X/3} M^{2/3}  \MP^{1/6}} e^{-i e \tau/ \ta(e)} \ .
\eeq
Integrating $d\tau$ and $d\omega$ in (\ref{tpdual1}) we obtain
 \bea
\label{tpdual1}
\tp(e) &=&\frac{m^{1/3}}{(2 \pi)^{1/2}e_{\rm I}^{1/2} 10^{X/3} M^{2/3}  \MP^{1/6}}  \int dt a(t)^3 f_{\omega=e/\ta(e)}^{-1}(t)\p(t) \\
&=& \frac{ \MP}{(2 \pi)^{1/2} m^{1/2} M e_{\rm I}^{1/2} }  \int dt \frac{1}{t}   e^{-i e t/ \ta(e) a(t) }  e^{i m t}
\eea
and the final result is  roughly
\beq
\label{dualphiI2}
\tp(e_{\rm I}) \simeq \frac{\MP }{m^{3/2}  M  t_{\rm now} e_{\rm I}^{1/2}} \ .
\eeq
The reciprocal of this at the inflation scale is
\beq
\label{dualphiI}
\p(t_{\rm I}) \simeq \frac{\MP M }{m^{3/2}  t_{\rm now} e_{\rm I}^{1/2}} \simeq \frac{m_{\rm now}^2 M_{\rm I}}{m^{3/2} \MP^{1/2}}\ .
\eeq
The  energy density  in the action $S$ (\ref{gravitydual}), evaluating only the contribution from the potential term, is 
\beq
m^2 |\p|^2 \simeq  \frac{ m_{\rm now}^4 M_{\rm I}^2 }{m  \MP } \ .
\eeq
The inflation energy density scale $ M_{\rm I}^4 $ is by far bigger than the one obtained here. 
A different method is  thus needed to explain the generation of the inflationary stage.

\section{Conclusion}
\label{conclusion}

With the technique of the generalised Fourier transform, we can realize Born reciprocity for every ordinary field theory, at least at the classical level.  The case considered in detail in this paper is the simplest one  possible:  one scalar matter field, gravity, and one abelian gauge field. This was enough to create a realistic cosmological scenario.
Our model does not  address the problem of the completion of gravity, or of what is the right description of gravity at the Planck scale. In this respect, it is just an effective description.

Our solution correlates the three following problems: the number of e-folds necessary to solve the horizon problem, the small value of the effective cosmological constant $\Lambda_{\rm  eff} =  10^{-120} \MP^4$, and the coincidence problem (i.e., why the cosmological constant becomes observable just now).   We are thus reducing the fine-tuning required in the ordinary $\Lambda$CDM model, although not all of the previous problems are solved. In our scenario inflation has to last just  the number of e-folds necessary to solve the horizon problem and this point remains with no justification. Another issue that has to be studied more in detail is the matching between the inflationary stage and the standard cosmological expansion.

\section*{Acknowledgments}
 This  work is funded by the grant ``Rientro dei Cervelli Rita Levi Montalcini'' of the Italian government.  This work is also supported by the INFN special research project, ``Gauge and String Theories'' (GAST).

\end{document}